\documentclass[preprint,aps,pra,showpacs,floatfix]{revtex4-1}
\usepackage{times}
\usepackage {euscript}
\usepackage{amsthm}
\usepackage{graphicx}
\usepackage{amsmath}
\usepackage{amssymb}
\usepackage{amsfonts}
\usepackage[english]{babel}
\usepackage{multirow,makecell,array}
\graphicspath{{figures/}}

\begin{document}
\thispagestyle{empty}

    \title{Finite nuclear size corrections to\\ the recoil effect in hydrogenlike ions}

\author{I.~A.~Aleksandrov$^{1, 2}$, A.~A.~Shchepetnov$^{1, 3}$, D.~A.~Glazov$^{1, 3}$, and V.~M.~Shabaev$^{1}$}

\affiliation{$^1$~Department of Physics, St. Petersburg State University, Ulianovskaya 1, Petrodvorets, 198504 Saint Petersburg, Russia\\ $^2$~ITMO University, Kronverkskii ave 49, 197101 Saint Petersburg, Russia\\ $^3$ State Scientific Centre ``Institute for Theoretical and Experimental Physics'' of National Research Centre ``Kurchatov Institute'', B. Cheremushkinskaya st. 25, 117218 Moscow, Russia
\vspace{10mm}
}

\begin{abstract}
The finite nuclear size corrections to the relativistic recoil effect in H-like ions are calculated within the Breit approximation. The calculations are performed for the $1s$, $2s$, and $2p_{1/2}$ states in the range $Z =$ 1--110. The obtained results are compared with previous evaluations of this effect. It is found that for heavy ions the previously neglected corrections amount to about 20\% of the total nuclear size contribution to the recoil effect calculated within the Breit approximation.

\end{abstract}

\maketitle
\section{Introduction}
For the last decade a great progress was achieved in experiments aimed at investigations of the finite nuclear size and nuclear recoil effects in highly charged ions. These effects lead to the isotope shifts of the binding energies which were measured in Refs.~\cite{Orts_2006, Brandau_2008}. In Ref.~\cite{Orts_2006}, the measurement was carried out at the electron beam ion trap (EBIT) using a high-resolution grating spectrometer. This experiment provided the first test of the relativistic theory of the nuclear recoil effect with highly charged ions (namely, B-like argon)~\cite{Tupitsyn_2003}. In Ref.~\cite{Brandau_2008}, the measurements of the isotope shifts in dielectronic recombination spectra for Li-like neodymium ions were used to determine the nuclear charge radii difference. The values obtained in this experiment were also sensitive to the relativistic nuclear recoil contribution (see Ref.~\cite{zubova} and references therein). It is expected that the accuracy of the isotope shift measurements will be significantly increased with the new FAIR facilities in Darmstadt~\cite{fair}, so that it is required to perform high precision calculations including the nuclear size corrections to the recoil effect.\\
\indent
It is well-known that in non-relativistic theory the nuclear recoil effect for a hydrogenlike atom can be easily taken into account to all orders in $m/M$ by using the reduced mass $\mu = mM/(m + M)$ instead of the electron mass $m$ ($M$ is the nuclear mass). The full relativistic theory of the nuclear recoil effect can be formulated only in the framework of QED~\cite{shabaev1985, shabaev1988, Yelkhovsky1994, pachucki_grotch1995, shabaev1998, adkins2007}. For the point-nucleus case, the total recoil correction of the first order in $m/M$ to the energy of a state $|a \rangle$ of a hydrogenlike ion can be written as a sum of a low-order term $\Delta E_\text{L}$ and a higher-order term $\Delta E_\text{H}$~\cite{shabaev1985, shabaev1988} (in units $\hbar = c = 1$):
\begin{eqnarray}
\Delta E &=& \Delta E_\text{L} + \Delta E_\text{H},\nonumber \\
\Delta E_\text{L} &=& \frac{1}{2M} \big \langle a \left | \boldsymbol{p}^2 - \big [ \boldsymbol{D}(0) \cdot \boldsymbol{p} + \boldsymbol{p} \cdot \boldsymbol{D} (0) \left. \big ] \right. \right | a \big \rangle, \label{eq:dE_tot_L}\\
\Delta E_\text{H} &=& \frac{i}{2\pi M} \int \limits_{-\infty}^\infty \mathrm{d}\omega \Bigg \langle a \left| \Bigg ( \right.\boldsymbol{D}(\omega) - \frac{[\boldsymbol{p}, V]}{\omega + i0} \Bigg ) G(\omega + E_a) \Bigg ( \boldsymbol{D}(\omega) + \frac{[\boldsymbol{p}, V]}{\omega + i0} \left. \Bigg ) \right| a \Bigg \rangle, \label{eq:dE_tot_H}
\end{eqnarray}
where $V(r) = -\alpha Z/r$ is the Coulomb potential of the nucleus, $G(\omega) = [\omega - H(1 - i0)]^{-1}$ is the relativistic Coulomb Green function, $H = \boldsymbol{\alpha} \cdot \boldsymbol{p} + \beta m + V$, $D_j (\omega, r) = -4\pi \alpha Z \alpha_i D_{ij} (\omega, r)$, and $D_{ij} (\omega, r)$ is the transverse part of the photon propagator in the Coulomb gauge which in coordinate space has the following form:
\begin{equation}
D_{ij}(\omega, r) = -\frac{1}{4\pi} \Bigg \{ \frac{\mathrm{exp}(i|\omega |r)}{r}\delta_{ij}  + \nabla_i \nabla_j \frac{\mathrm{exp}(i|\omega |r) - 1}{\omega^2 r} \Bigg \}.
\label{eq:propagator}
\end{equation}
The first term $\Delta E_\text{L}$ contains all the recoil corrections within the $(\alpha Z)^4m^2/M$ approximation (the so-called Breit approximation). Its calculation, based on the virial relations for the Dirac equation~\cite{epstein1962, shabaev_virial}, leads to~\cite{shabaev1985}
\begin{equation}
\Delta E_\text{L} = \frac{m^2 - E_{a0}^2}{2M},
\label{eq:dE_L_point}
\end{equation}
where $E_{a0}$ is the Dirac electron energy for the point-nucleus case. The second term $\Delta E_\text{H}$ contains the contribution of order $(\alpha Z)^5m^2/M$ and all contributions of higher orders in $\alpha Z$ which are not included in $\Delta E_\text{L}$. Its evaluation to all orders in $\alpha Z$ was performed in Refs.~\cite{artemyev1995, artemyev1995_2, adkins2007}.\\
\indent
According to Ref.~\cite{shabaev1998}, the nuclear size corrections to the recoil effect can be partly taken into account by employing the potential of an extended nucleus in the formulas~(\ref{eq:dE_tot_L}) and (\ref{eq:dE_tot_H}) including the values of $E_a, |a \rangle$, and $G(\omega)$. The corresponding calculations were carried out in Refs.~\cite{shabaev_recoil, shabaev_recoil2}. This approach allows one to evaluate the nuclear size corrections completely for the Coulomb part of the recoil effect
\begin{equation}
\Delta E_\text{C} = \Big \langle a \left | \left. \frac{\boldsymbol{p}^2}{2M} \right. \right | a \Big \rangle + \frac{2\pi i}{M} \int \limits_{-\infty}^\infty \mathrm{d}\omega \delta_{+}^2 (\omega) \langle a | [\boldsymbol{p}, V] G(\omega + E_a) [\boldsymbol{p}, V] | a \rangle
\label{eq:dE_C}
\end{equation}
and only partly for the one-transverse-photon and two-transverse-photon parts:
\begin{eqnarray}
\Delta E_\text{tr(1)} &=&  - \frac{1}{2M} \langle a | \boldsymbol{D}(0) \boldsymbol{p} + \boldsymbol{p} \boldsymbol{D}(0) | a \rangle \nonumber\\
&& - \frac{1}{M} \int \limits_{-\infty}^\infty \mathrm{d}\omega \delta_{+} (\omega) \langle a | [\boldsymbol{p}, V] G(\omega + E_a)\boldsymbol{D}(\omega) - \boldsymbol{D}(\omega) G(\omega + E_a)  [\boldsymbol{p}, V] | a \rangle, \label{eq:dE_tr1}\\
\Delta E_\text{tr(2)} &=& \frac{i}{2\pi M} \int \limits_{-\infty}^\infty \mathrm{d}\omega \langle a | \boldsymbol{D}(\omega) G(\omega + E_a) \boldsymbol{D}(\omega) | a \rangle. \label{eq:dE_tr2}
\end{eqnarray}
In this paper we present the complete evaluation of the nuclear size correction to the low-order nuclear recoil contribution $\Delta E_\text{L}$ which corresponds to the Breit approximation. The calculations are performed for the $1s$, $2s$, and $2p_{1/2}$ states in the range $Z =$ 1--110.
\section{Nuclear recoil operator within the Breit approximation}
\begin{figure} [h] 
  \center
  \includegraphics [height = 5 cm] {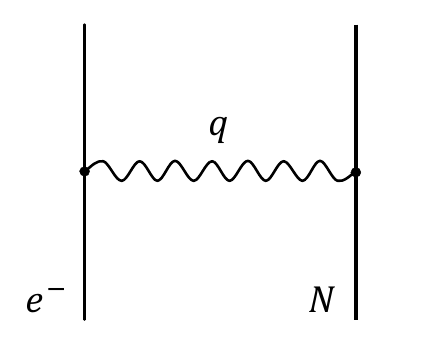}
  \caption{One-photon exchange diagram.} 
  \label{fig:one_photon}
\end{figure}

To derive the nuclear recoil operator for a hydrogenlike ion within the Breit approximation, one should account for the one-photon exchange between the electron and the nucleus in the Coulomb gauge (see Fig.~\ref{fig:one_photon}) and consider the nucleus as a non-relativistic particle. Discarding the nucleus spin-dependent terms, one gets the following electron-nucleus interaction potential in momentum space~\cite{grotch, borie82}:
\begin{equation}
V_\text{eff}(\boldsymbol{q}) = -\alpha Z \Bigg ( \frac{F(\boldsymbol{q})}{\boldsymbol{q}^2} + \frac{1}{2M}\left\{ \frac{F(\boldsymbol{q})}{\boldsymbol{q}^2}, \boldsymbol{\alpha} \cdot \boldsymbol{p} \right\} - \frac{1}{2M}\bigg [ \boldsymbol{\alpha} \cdot \boldsymbol{p}, \bigg [ \boldsymbol{p}^2, \frac{F(\boldsymbol{q})}{\boldsymbol{q}^4} \bigg ] \bigg ]\Bigg ),
\label{eq:Veff_mom}
\end{equation}
where $F(\boldsymbol{q})$ is the nuclear form factor. In coordinate space Eq.~(\ref{eq:Veff_mom}) reads:
\begin{equation}
V_\text{eff}(r) = V(r) + \frac{1}{2M}\left\{ V(r), \boldsymbol{\alpha} \cdot \boldsymbol{p} \right\} + \frac{1}{4M}\big [ \boldsymbol{\alpha} \cdot \boldsymbol{p}, \big [ \boldsymbol{p}^2, W(r) \big ] \big ],
\label{eq:Veff_pos}
\end{equation}
where
\begin{eqnarray}
V(r) &=& -\alpha Z\int \mathrm{d}\boldsymbol{r}^\prime \frac{\rho(\boldsymbol{r}^\prime)}{|\boldsymbol{r} - \boldsymbol{r}^\prime|}, \label{eq:V}\\
W(r) &=& -\alpha Z\int \mathrm{d}\boldsymbol{r}^\prime \rho(\boldsymbol{r}^\prime)|\boldsymbol{r} - \boldsymbol{r}^\prime|, \label{eq:W}
\end{eqnarray}
and $\rho (\boldsymbol{r})$ is the density of the nuclear charge distribution $\big ( \int \mathrm{d}\boldsymbol{r} \rho (\boldsymbol{r}) = 1\big )$. Taking into account the nonrelativistic kinetic energy of the nucleus in the centre-of-mass frame, one obtains the low-order nuclear recoil operator accounting for the nuclear size effect:
\begin{equation}
H_\text{M} = \frac{\boldsymbol{p}^2}{2M} + \frac{1}{2M}\left\{ V(r), \boldsymbol{\alpha} \cdot \boldsymbol{p} \right\} + \frac{1}{4M}\big [ \boldsymbol{\alpha} \cdot \boldsymbol{p}, \big [ \boldsymbol{p}^2, W(r) \big ] \big ].
\label{eq:H_M}
\end{equation}
To first order in $m/M$, the nuclear recoil contribution is given by the expectation value of $H_\text{M}$~\cite{borie82}:
\begin{equation}
\Delta E = \langle a | H_\text{M} | a \rangle = \frac{1}{2M}(E_{a}^2 - m^2) - \frac{m}{M} \langle a | \beta V(r) | a \rangle - \frac{1}{2M} \langle a | W^\prime (r) V^\prime(r) + V^2(r) | a \rangle.
\label{eq:dE_final}
\end{equation}
For the case of a point nucleus $W^\prime (r) V^\prime(r) + V^2(r) = 0$ and, using the virial relations~\cite{epstein1962, shabaev_virial}, we get Eq.~(\ref{eq:dE_L_point}). \\
\indent
To evaluate the finite nuclear size corrections to the recoil effect expressed by Eq.~(\ref{eq:dE_final}) (it corresponds to the low-order term $\Delta E_\text{L}$) one should use the values of $E_a$, $|a \rangle$, $V(r)$, and $W(r)$ for an extended nucleus. For a spherically symmetric $\rho (\boldsymbol{r}) = \rho (r)$ we have
\begin{eqnarray}
V(r) &=& -4\pi \alpha Z\bigg [ \frac{1}{r} \int \limits_{0}^r \mathrm{d} r^\prime r^{\prime 2} \rho(r^\prime) + \int \limits_{r}^\infty \mathrm{d} r^\prime r^\prime \rho(r^\prime) \bigg ], \label{eq:V_r}\\
W(r) &=& -4\pi \alpha Z\bigg [ \frac{1}{r} \int \limits_{0}^r \mathrm{d} r^\prime r^{\prime 2}\Big ( r^2 + \frac{r^{\prime 2}}{3} \Big ) \rho(r^\prime) + \int \limits_{r}^\infty \mathrm{d} r^\prime r^\prime \Big ( r^{\prime2} + \frac{r^2}{3} \Big ) \rho(r^\prime) \bigg ]. \label{eq:W_r}
\end{eqnarray}
Assuming the nucleus to be a homogeneously charged sphere with radius $R = \sqrt{5/3}\, \langle r^2 \rangle^{1/2}$ one can obtain
\begin{eqnarray}
V (r) &=&
\begin{cases}
-\frac{\alpha Z}{r} & r \geq R\\
-\frac{\alpha Z}{R} \Big( \frac{3}{2} - \frac{r^2}{2R^2} \Big) & r<R\\
\end{cases},\\
W (r) &=&
\begin{cases}
-\alpha Z r \Big ( 1 + \frac{R^2}{5r^2} \Big) & r \geq R\\
-\alpha Z R \Big ( \frac{3}{4} + \frac{r^2}{2R^2} - \frac{r^4}{20R^4} \Big ) & r<R\\
\end{cases}.
\end{eqnarray}
The nuclear size corrections to the low-order recoil term were calculated numerically and the corresponding results are presented in the next section.
\section{Numerical results and discussion}
The nuclear size corrections to the low-order recoil effect for the $1s$, $2s$, and $2p_{1/2}$ states are summarized in Tables~\ref{table:res_1s}, \ref{table:res_2s}, and \ref{table:res_2p}, respectively. They are expressed in terms of the function $\Delta F_\text{L} (\alpha Z)$ which is defined by
\begin{equation}
\Delta E_\text{L} = \frac{m^2 - E_{a0}^2}{2M} \big [ 1 + \Delta F_\text{L} (\alpha Z) \big].
\label{eq:dE_F}
\end{equation}
The homogeneously-charged-sphere model was used to describe the nuclear charge distribution. The function $\Delta F_\text{L}^\text{(point)} (\alpha Z)$ corresponds to the approximate evaluation of Refs.~\cite{shabaev_recoil, shabaev_recoil2} that employs the point-nucleus recoil operator and the extended-nucleus Dirac wave functions. Our calculations within this approximation are in perfect agreement with those from Refs.~\cite{shabaev_recoil, shabaev_recoil2}. The function $\Delta F_\text{L}^\text{(finite)} (\alpha Z)$ represents the values obtained by using Eq.~(\ref{eq:dE_final}) and includes all finite nuclear size corrections to the Breit term $\Delta E_\text{L}$. The difference between $\Delta F_\text{L}^\text{(finite)} (\alpha Z)$ and $\Delta F_\text{L}^\text{(point)} (\alpha Z)$ is displayed in the fifth column \big [$\delta \Delta F_\text{L} (\alpha Z) = \Delta F_\text{L}^\text{(finite)} (\alpha Z) - \Delta F_\text{L}^\text{(point)} (\alpha Z)$\big ].
\begin{table} [t]
\centering
\begin{ruledtabular}
\begin{tabular}{rcllll}
\multicolumn{1}{c}{\raisebox{0pt}[12pt][5pt]{$Z$}} &
\multicolumn{1}{c}{\raisebox{0pt}[12pt][5pt]{$\langle r^2 \rangle^{1/2}$~[fm]}} &
\multicolumn{1}{c}{\raisebox{0pt}[12pt][5pt]{$\Delta F_\text{L}^\text{(point)} (\alpha Z)$}} &
\multicolumn{1}{c}{\raisebox{0pt}[12pt][5pt]{$\Delta F_\text{L}^\text{(finite)} (\alpha Z)$}} &
\multicolumn{1}{c}{\raisebox{0pt}[12pt][5pt]{$\delta \Delta F_\text{L} (\alpha Z)$}} &
\multicolumn{1}{c}{\raisebox{0pt}[12pt][5pt]{$\Delta F_\text{H}^\text{(point)} (\alpha Z)$~\cite{shabaev_recoil}}}\\
\hline
\raisebox{0pt}[10pt][4pt]{1} &
\raisebox{0pt}[10pt][4pt]{$0.878$} &
\raisebox{0pt}[10pt][4pt]{$-0.345\times 10^{-8}$} &
\raisebox{0pt}[10pt][4pt]{$-0.345\times 10^{-8}$} &
\multicolumn{1}{c}{\raisebox{0pt}[10pt][4pt]{}} &
\raisebox{0pt}[10pt][4pt]{$0.23\times 10^{-8}$}\\
2 & $1.676$ & $-0.520\times 10^{-7}$ & $-0.519\times 10^{-7}$ & \multicolumn{1}{c}{} & $0.35\times 10^{-7}$ \\
5 & $2.406$ & $-0.102\times 10^{-5}$ & $-0.102\times 10^{-5}$ & \multicolumn{1}{c}{} & $0.76\times 10^{-6}$ \\
10 & $3.006$ & $-0.969\times 10^{-5}$ & $-0.964\times 10^{-5}$ & $0.5\times 10^{-7}$ & $0.77\times 10^{-5}$ \\
20 & $3.478$ & $-0.934\times 10^{-4}$ & $-0.920\times 10^{-4}$ & $0.14\times 10^{-5}$ & $0.71\times 10^{-4}$ \\
30 & $3.949$ & $-0.408\times 10^{-3}$ & $-0.396\times 10^{-3}$ & $0.12\times 10^{-4}$ & $0.29\times 10^{-3}$ \\
40 & $4.285$ & $-0.127\times 10^{-2}$ & $-0.121\times 10^{-2}$ & $0.6\times 10^{-4}$ & $0.79\times 10^{-3}$ \\
50 & $4.644$ & $-0.339\times 10^{-2}$ & $-0.316\times 10^{-2}$ & $0.23\times 10^{-3}$ & $0.19\times 10^{-2}$ \\
60 & $4.942$ & $-0.826\times 10^{-2}$ & $-0.752\times 10^{-2}$ & $0.74\times 10^{-3}$ & $0.35\times 10^{-2}$ \\
70 & $5.305$ & $-0.0194$ & $-0.0172$ & $0.22\times 10^{-2}$ & $0.52\times 10^{-2}$ \\
80 & $5.458$ & $-0.0436$ & $-0.0376$ & $0.60\times 10^{-2}$ & $0.17\times 10^{-2}$ \\
90 & $5.785$ & $-0.0992$ & $-0.0830$ & $0.0162$ & $-0.034$ \\
92 & $5.857$ & $-0.117$ & $-0.097$ & $0.020$ & $-0.053$ \\
100 & $5.886$ & $-0.224$ & $-0.182$ & $0.042$ & $-0.25$ \\
110 & $5.961$ & $-0.517$ & $-0.407$ & $0.110$ & $-1.7$ \\
\end{tabular}
\end{ruledtabular}
\caption{Finite nuclear size corrections to the low-order recoil contribution for the $1s$ state expressed in terms of the function $\Delta F_\text{L} (\alpha Z)$, which is defined by Eq.~(\ref{eq:dE_F}). The values of the nuclear radii used were taken from Ref.~\cite{angeli2013} (for $Z \leq 92$) and Ref.~\cite{johnson_soff} (for $Z = 100$,~$110$). The last column refers to the higher-order recoil term which was evaluated in Ref.~\cite{shabaev_recoil} using slightly different values of the nuclear radii.}
\label{table:res_1s}
\end{table}
\begin{table} [h]
\centering
\begin{ruledtabular}
\begin{tabular}{rcllll}
\multicolumn{1}{c}{\raisebox{0pt}[12pt][5pt]{$Z$}} &
\multicolumn{1}{c}{\raisebox{0pt}[12pt][5pt]{$\langle r^2 \rangle^{1/2}$~[fm]}} &
\multicolumn{1}{c}{\raisebox{0pt}[12pt][5pt]{$\Delta F_\text{L}^\text{(point)} (\alpha Z)$}} &
\multicolumn{1}{c}{\raisebox{0pt}[12pt][5pt]{$\Delta F_\text{L}^\text{(finite)} (\alpha Z)$}} &
\multicolumn{1}{c}{\raisebox{0pt}[12pt][5pt]{$\delta \Delta F_\text{L} (\alpha Z)$}} &
\multicolumn{1}{c}{\raisebox{0pt}[12pt][5pt]{$\Delta F_\text{H}^\text{(point)} (\alpha Z)$~\cite{shabaev_recoil2}}}\\
\hline
\raisebox{0pt}[10pt][4pt]{1} &
\raisebox{0pt}[10pt][4pt]{$0.878$} &
\raisebox{0pt}[10pt][4pt]{$-0.172\times 10^{-8}$} &
\raisebox{0pt}[10pt][4pt]{$-0.172\times 10^{-8}$} &
\multicolumn{1}{c}{\raisebox{0pt}[10pt][4pt]{}} &
\multicolumn{1}{c}{\raisebox{0pt}[10pt][4pt]{}}\\
2 & $1.676$ & $-0.260\times 10^{-7}$ & $-0.260\times 10^{-7}$ & \multicolumn{1}{c}{} & \multicolumn{1}{c}{} \\
5 & $2.406$ & $-0.513\times 10^{-6}$ & $-0.513\times 10^{-6}$ & \multicolumn{1}{c}{} & \multicolumn{1}{c}{} \\
10 & $3.006$ & $-0.487\times 10^{-5}$ & $-0.484\times 10^{-5}$ & $0.3\times 10^{-7}$ & $0.4\times 10^{-5}$ \\
20 & $3.478$ & $-0.474\times 10^{-4}$ & $-0.466\times 10^{-4}$ & $0.8\times 10^{-6}$ & $0.4\times 10^{-4}$ \\
30 & $3.949$ & $-0.211\times 10^{-3}$ & $-0.204\times 10^{-3}$ & $0.7\times 10^{-5}$ & $0.1\times 10^{-3}$ \\
40 & $4.285$ & $-0.669\times 10^{-3}$ & $-0.637\times 10^{-3}$ & $0.32\times 10^{-4}$ & $0.4\times 10^{-3}$ \\
50 & $4.644$ & $-0.185\times 10^{-2}$ & $-0.172\times 10^{-2}$ & $0.13\times 10^{-3}$ & $0.1\times 10^{-2}$ \\
60 & $4.942$ & $-0.468\times 10^{-2}$ & $-0.426\times 10^{-2}$ & $0.42\times 10^{-3}$ & $0.2\times 10^{-2}$ \\
70 & $5.305$ & $-0.0115$ & $-0.0102$ & $0.13\times 10^{-2}$ & $0.3\times 10^{-2}$ \\
80 & $5.458$ & $-0.0271$ & $-0.0233$ & $0.0038$ & $0.006$ \\
90 & $5.785$ & $-0.0653$ & $-0.0545$ & $0.0108$ & \multicolumn{1}{c}{} \\
92 & $5.857$ & $-0.0779$ & $-0.0646$ & $0.0133$ & $-0.03$ \\
100 & $5.886$ & $-0.156$ & $-0.127$ & $0.029$ & \multicolumn{1}{c}{} \\
110 & $5.961$ & $-0.382$ & $-0.300$ & $0.082$ & \multicolumn{1}{c}{}\\
\end{tabular}
\end{ruledtabular}
\caption{Finite nuclear size corrections to the low-order recoil contribution for the $2s$ state expressed in terms of the function $\Delta F_\text{L} (\alpha Z)$, which is defined by Eq.~(\ref{eq:dE_F}). The values of the nuclear radii used were taken from Ref.~\cite{angeli2013} (for $Z \leq 92$) and Ref.~\cite{johnson_soff} (for $Z = 100$,~$110$). The last column refers to the higher-order recoil term which was evaluated in Ref.~\cite{shabaev_recoil2} using slightly different values of the nuclear radii.}
\label{table:res_2s}
\end{table}
\begin{table} [h]
\centering
\begin{ruledtabular}
\begin{tabular}{rclll}
\multicolumn{1}{c}{\raisebox{0pt}[12pt][5pt]{$Z$}} &
\multicolumn{1}{c}{\raisebox{0pt}[12pt][5pt]{$\langle r^2 \rangle^{1/2}$~[fm]}} &
\multicolumn{1}{c}{\raisebox{0pt}[12pt][5pt]{$\Delta F_\text{L}^\text{(point)} (\alpha Z)$}} &
\multicolumn{1}{c}{\raisebox{0pt}[12pt][5pt]{$\Delta F_\text{L}^\text{(finite)} (\alpha Z)$}} &
\multicolumn{1}{c}{\raisebox{0pt}[12pt][5pt]{$\delta \Delta F_\text{L} (\alpha Z)$}}\\
\hline
\raisebox{0pt}[10pt][4pt]{10} &
\raisebox{0pt}[10pt][4pt]{$3.006$} &
\raisebox{0pt}[10pt][4pt]{$-0.942\times 10^{-8}$} &
\raisebox{0pt}[10pt][4pt]{$-0.905\times 10^{-8}$} &
\raisebox{0pt}[10pt][4pt]{$0.37\times 10^{-9}$} \\
20 & $3.478$ & $-0.207\times 10^{-6}$ & $-0.196\times 10^{-6}$ & $0.11\times 10^{-7}$ \\
30 & $3.949$ & $-0.204\times 10^{-5}$ & $-0.192\times 10^{-5}$ & $0.12\times 10^{-6}$ \\
40 & $4.285$ & $-0.117\times 10^{-4}$ & $-0.109\times 10^{-4}$ & $0.8\times 10^{-6}$ \\
50 & $4.644$ & $-0.518\times 10^{-4}$ & $-0.474\times 10^{-4}$ & $0.44\times 10^{-5}$ \\
60 & $4.942$ & $-0.196\times 10^{-3}$ & $-0.175\times 10^{-3}$ & $0.21\times 10^{-4}$ \\
70 & $5.305$ & $-0.688\times 10^{-3}$ & $-0.600\times 10^{-3}$ & $0.88\times 10^{-4}$ \\
80 & $5.458$ & $-0.225\times 10^{-2}$ & $-0.191\times 10^{-2}$ & $0.34\times 10^{-3}$ \\
90 & $5.785$ & $-0.741\times 10^{-2}$ & $-0.613\times 10^{-2}$ & $0.128\times 10^{-2}$ \\
92 & $5.857$ & $-0.942\times 10^{-2}$ & $-0.774\times 10^{-2}$ & $0.168\times 10^{-2}$ \\
100 & $5.886$ & $-0.0242$ & $-0.0195$ & $0.0047$ \\
110 & $5.961$ & $-0.0825$ & $-0.0643$ & $0.0182$ \\
\end{tabular}
\end{ruledtabular}
\caption{Finite nuclear size corrections to the low-order recoil contribution for the $2p_{1/2}$ state expressed in terms of the function $\Delta F_\text{L} (\alpha Z)$, which is defined by Eq.~(\ref{eq:dE_F}). The values of the nuclear radii used were taken from Ref.~\cite{angeli2013} (for $Z \leq 92$) and Ref.~\cite{johnson_soff} (for $Z = 100$,~$110$).}
\label{table:res_2p}
\end{table}
The finite nuclear size corrections to the higher-order recoil term $\Delta E_\text{H}$, calculated using the point-nucleus recoil operator~\cite{shabaev_recoil, shabaev_recoil2}, are also presented. As was shown in Ref.~\cite{shabaev_recoil}, the leading nuclear size corrections to the low-order and higher-order terms cancel each other for $\alpha Z \ll 1$. The difference between $\Delta F_\text{L}^\text{(finite)} (\alpha Z)$ and $\Delta F_\text{L}^\text{(point)} (\alpha Z)$, being negligible for low-Z ions, grows when $Z$ increases and reaches about 20~\% of $\Delta F_\text{L}^\text{(point)} (\alpha Z)$ at $Z = 110$. It is also worth noting that the values of $\Delta F_\text{H}^\text{(point)} (\alpha Z)$ change the sign at $Z \approx 80$, leading to a strong enhancement of the total nuclear size correction for heavy ions.\\
\indent
Concluding, in the present paper we have performed the complete evaluation of the nuclear size correction to the low-order (Breit) recoil contribution. We have found that the corrections which are beyond the previously used approximation~\cite{shabaev_recoil, shabaev_recoil2} can contribute on the level of~$20\%$ of the total nuclear size contribution to the low-order recoil effect. To calculate the nuclear size corrections to the higher-order recoil term, which are beyond the approximation used in Refs.~\cite{shabaev_recoil, shabaev_recoil2}, one should first derive the corresponding corrections to formulas~(\ref{eq:dE_tr1})--(\ref{eq:dE_tr2}). Such a derivation, which seems rather problematic, requires further theoretical investigations that are beyond the scope of this paper.
\section*{Acknowledgements}
This work was supported by~RFBR (Grant No.~13-02-00630), by~SPbSU (Grant No.~11.38.269.2014), and by~the~FAIR--Russia Research Center. I.~A.~A. acknowledges the financial support by the Dynasty foundation.


\begin{thebibliography}{99}
%
\bibitem{Orts_2006} R.~Soria Orts, Z.~Harman, J.~R.~Crespo Lopez-Urrutia, A.~N.~Artemyev, H.~Bruhns, A.~J.~Gonzalez Martinez, U.~D.~Jentschura, C.~H.~Keitel, A.~Lapierre, V.~Mironov, V.~M.~Shabaev, H.~Tawara, I.~I.~Tupitsyn, J.~Ullrich, and A.~V.~Volotka, Phys. Rev. Lett. {\bf 97}, 103002 (2006).
%
\bibitem{Brandau_2008} C.~Brandau, C.~Kozhuharov, Z.~Harman, A.~M\"uller,  S.~Schippers,  Y.~S.~Kozhedub,  D.~Bernhardt, S.~B\"ohm, J.~Jacobi, E.~W.~Schmidt, P.~H.~Mokler, F.~Bosch, H.-J.~Kluge, Th.~St\"ohlker, K.~Beckert, P.~Beller, F.~Nolden, M.~Steck, A.~Gumberidze, R.~Reuschl, U.~Spillmann, F.~J.~Currell, I.~I.~Tupitsyn, V.~M.~Shabaev, U.~D.~Jentschura, C.~H.~Keitel, A.~Wolf, and Z.~Stachura, Phys. Rev. Lett. {\bf 100}, 073201 (2008).
%
\bibitem{Tupitsyn_2003} I.~I.~Tupitsyn, V.~M.~Shabaev, J.~R.~Crespo Lopez-Urrutia, I.~Draganic, R.~Soria Orts, and J.~Ullrich, Phys. Rev.~A {\bf 68}, 022511 (2003).
%
\bibitem{zubova} N.~A.~Zubova, Y.~S.~Kozhedub, V.~M.~Shabaev, I.~I.~Tupitsyn, A.~V.~Volotka, G.~Plunien, C.~Brandau, and Th.~St\"ohlker, arXiv 1410.7071.
%
\bibitem{fair} http://www.fair-center.eu
%
\bibitem{shabaev1985} V.~M.~Shabaev, Teor. Mat. Fiz. {\bf 63}, 394 (1985) \big [Theor. Math. Phys. {\bf 63}, 588 (1985)\big ].
%
\bibitem{shabaev1988} V.~M.~Shabaev, Yad. Fiz. {\bf 47}, 107 (1988) \big [Sov. J. Nucl. Phys. {\bf 47}, 69 (1988)\big ].
%
\bibitem{Yelkhovsky1994} A.~S.~Yelkhovsky, Budker Institute of Nuclear Physics, Novosibirsk, Report No. BINP 94-27, hep-th/9403095 (1994).
%
\bibitem{pachucki_grotch1995} K.~Pachucki and H.~Grotch, Phys. Rev.~A {\bf 51}, 1854 (1995).
%
\bibitem{shabaev1998} V.~M.~Shabaev, Phys. Rev.~A {\bf 57}, 59 (1998).
%
\bibitem{adkins2007} G.~S.~Adkins, S.~Morrison, and J.~Sapirstein, Phys. Rev.~A {\bf 76}, 042508 (2007).
%
\bibitem{artemyev1995} A.~N.~Artemyev, V.~M.~Shabaev, and V.~A.~Yerokhin, Phys. Rev.~A {\bf 52}, 1884 (1995).
%
\bibitem{artemyev1995_2} A.~N.~Artemyev, V.~M.~Shabaev, and V.~A.~Yerokhin, J. Phys.~B {\bf 28}, 5201 (1995).
%
\bibitem{shabaev_recoil} V.~M.~Shabaev, A.~N.~Artemyev, T.~Beier, G.~Plunien, V.~A.~Yerokhin, and G.~Soff, Phys. Rev.~A {\bf 57}, 4235 (1998).
%
\bibitem{shabaev_recoil2} V.~M.~Shabaev, A.~N.~Artemyev, T.~Beier, G.~Plunien, V.~A.~Yerokhin, and G.~Soff, Phys. Scr. T {\bf 80}, 493 (1999).
%
\bibitem{epstein1962} J.~Epstein and S.~Epstein, Am.~J. Phys. {\bf 30}, 266 (1962).
%
\bibitem{shabaev_virial} V.~M.~Shabaev, J. Phys. B {\bf 24}, 4479 (1991).
%
\bibitem{grotch} H.~Grotch and D.~R.~Yennie, Rev. Mod. Phys. {\bf 41},  350 (1969).
%
\bibitem{borie82} E.~Borie and G.~A.~Rinker, Rev. Mod. Phys. {\bf 54}, 67 (1982).
%
\bibitem{angeli2013} I.~Angeli and K.~P.~Marinova, At. Data Nucl. Data Tables {\bf 99}, 69 (2013).
%
\bibitem{johnson_soff} W.~R.~Johnson and G.~Soff, At. Data Nucl. Data Tables {\bf 33}, 406 (1985).
%
\end{thebibliography}
\end{document}